\documentclass[reprint,nofootinbib,showpacs]{revtex4}
\usepackage{graphicx}  
\usepackage{dcolumn}   
\usepackage{bm}        
\usepackage{amssymb}
\usepackage{hyperref}
\usepackage{multirow}

\begin{document}

\title{Signature of lepton flavor universality violation
in $B_s \to D_s \tau \nu$ semileptonic decays}

\author{Rupak~Dutta${}^{}$}
\email{rupak@phy.nits.ac.in}
\author{N Rajeev${}^{}$}
\email{rajeev@rs.phy.student.nits.ac.in}
\affiliation{
${}$National Institute of Technology Silchar, Silchar 788010, India\\
}

\begin{abstract}
Deviation from the standard model prediction is observed in many semileptonic $B$ decays mediated via $b \to c$ charged current 
interactions. In particular, current experimental measurements of the ratio of branching ratio $R_D$ and $R_{D^{\ast}}$ in 
$B \rightarrow D^{(*)}l \nu$ decays disagree with standard model expectations at the level of about $4.1\sigma$. Moreover, recent 
measurement of the ratio of
branching ratio $R_{J/\Psi}$ by LHCb, where $R_{J/\Psi} = \mathcal B(B_c \to J/\Psi\,\tau\nu)/\mathcal B(B_c \to J/\Psi\,\mu\nu)$, is more
than $2\sigma$ away from the standard model prediction. In this context,
we consider an effective Lagrangian in the presence of vector and scalar new physics couplings to study the 
implications of $R_D$ and $R_{D^{\ast}}$ anomalies in $B_s \to D_s\,\tau\nu$ decays.  
We give prediction of several observables such as branching ratio, ratio of branching ratio, forward backward asymmetry parameter, $\tau$
polarization fraction, and the convexity parameter for the $B_s \to D_s\,\tau\nu$ decays within the standard model and within various
new physics scenarios.
\end{abstract}

\maketitle

\section{Introduction}
Anomalies present in $R_D$, $R_{D^*}$ and $R_{J/\Psi}$ challenged the lepton flavor universality. 
At present the deviation in $R_D$ and $R_{D^*}$ from the standard model (SM) expectation~\cite{Lattice:2015rga,Na:2015kha,Aoki:2016frl,Bigi:2016mdz,Fajfer:2012vx,Bigi:2017jbd} 
is at the level of $4.1\sigma$~\cite{Lees:2013uzd,Huschle:2015rga,Sato:2016svk,Hirose:2016wfn,Aaij:2015yra,Amhis:2016xyh}. 
A similar deviation of $1.3\sigma$ has been reported by LHCb in the ratio $R_{J/\Psi}$~\cite{Aaij:2017tyk} as well.
Inspired by these anomalies we study the corresponding $B_s \to D_s \tau \nu$ decay mode within the SM and within various new physics (NP) scenarios
by using the $B_s$ to $D_s$ transition form factors obtained in lattice QCD of Ref.~\cite{Monahan:2017uby}. 
The $B_s \to D_s \tau \nu$ decays serve as a complementary decay channel to similar $B$ decays mediated 
via $b \to cl\nu$ quark level transition. Again, in the limit of $SU(3)$ flavor symmetry $B \to D l \nu$ and $B_s \to D_s l \nu$ decay modes 
should exhibit the similar properties.

Our main motivation here is to study the implication of $R_D$ and $R_{D^*}$
anomalies on $B_s \to D_s \tau \nu$ decay mode in a model independent way. We use the effective Lagrangian in the presence of NP couplings
and give the predictions of various physical observables.

\section{Theory}
The effective Lagrangian for $b \to cl\nu$ quark level transition decays consisting of the SM and beyond SM operators is given 
by~\cite{Cirigliano:2009wk,Bhattacharya:2011qm} \\
\begin{eqnarray}
\label{effl}
\mathcal L_{\rm eff} &=&
-\frac{4\,G_F}{\sqrt{2}}\,V_{c b}\,\Bigg\{(1 + V_L)\,\bar{l}_L\,\gamma_{\mu}\,\nu_L\,\bar{c}_L\,\gamma^{\mu}\,b_L +
V_R\,\bar{l}_L\,\gamma_{\mu}\,\nu_L\,\bar{c}_R\,\gamma^{\mu}\,b_R \nonumber \\
&&+
\widetilde{V}_L\,\bar{l}_R\,\gamma_{\mu}\,\nu_R\,\bar{c}_L\,\gamma^{\mu}\,b_L 
+
\widetilde{V}_R\,\bar{l}_R\,\gamma_{\mu}\,\nu_R\,\bar{c}_R\,\gamma^{\mu}\,b_R +
S_L\,\bar{l}_R\,\nu_L\,\bar{c}_R\,b_L \nonumber \\
&&+
S_R\,\bar{l}_R\,\nu_L\,\bar{c}_L\,b_R +
\widetilde{S}_L\,\bar{l}_L\,\nu_R\,\bar{c}_R\,b_L +
\widetilde{S}_R\,\bar{l}_L\,\nu_R\,\bar{c}_L\,b_R\Bigg\} + {\rm h.c.}\,,
\end{eqnarray}
where, $G_F$ is the Fermi coupling constant and $|V_{cb}|$ is the CKM matrix element and the couplings such as
$V_L$, $V_R$, $S_L$, $S_R$ and $\tilde{V}_L$, $\tilde{V}_R$, $\tilde{S}_L$, $\tilde{S}_R$ denote the NP Wilson 
coefficients involving left handed and right handed neutrinos respectively. 
We investigate several $q^2$ dependent observables such as
differential branching ratio DBR $(q^2)$, ratio of branching ratio $R(q^2)$, lepton side forward backward asymmetry $A_{FB}^l (q^2)$, 
polarization fraction of the charged lepton $P_l (q^2)$, and convexity parameter $C_{F}^l (q^2)$ defined as,
\begin{eqnarray}
{DBR}(q^2)=\frac{d\Gamma/dq^2}{\Gamma_{\rm Tot}}\,, \qquad 
A_{FB}(q^2) = \frac{\Big(\int_{-1}^{0}-\int_{0}^{1}\Big)d\cos\theta\frac{d\Gamma}{dq^2\,d\cos\theta}}{\frac{d\Gamma}{dq^2}} \,, \nonumber  \\ 
R(q^2)=\frac{\mathcal B(B_s \rightarrow D_s \tau \nu)}{\mathcal B(B_s \rightarrow D_s\,l\,\nu)}\,, \qquad \nonumber 
P_{l}(q^2)=\frac{d\Gamma(+)/dq^2 - d\Gamma(-)/dq^2}{d\Gamma(+)/dq^2 + d\Gamma(-)/dq^2}, \nonumber \\
C_{F}^{l}(q^2)= \frac{1}{\left(d\Gamma/dq^2\right)} \left(\frac{d}{d\cos\theta}\right)^2 \left[\frac{d\Gamma}{dq^2\,d\cos\theta}\right]
\end{eqnarray}

We also give predictions on the average values of these observable by separately integrating the numerator and the denominator over $q^2$.
A detailed discussion is reported in the Ref.~\cite{Dutta:2018jxz}.

\section{Results and Discussions}
We report in Table~\ref{tabsm} the SM central values and the corresponding ranges of each observable for $B_s \to D_s l \nu$ decay mode.
The SM central values are obtained by considering the central values of each input parameters and the ranges are obtained 
by including the uncertainties associated with $|V_{cb}|$ and the form factor inputs.
The details are presented in the Ref.~\cite{Dutta:2018jxz}.
We notice that the branching ratio of $B_s \to D_s l \nu$ is of the order of $10^{-2}$ for the $e$ mode and the $\tau$ mode. 
In SM, the observables $A_{FB}^l$, $P_l$, $C_{F}^l$ for the $e$ mode are observed to be quite different from the corresponding $\tau$ mode. 

\begin{figure}[htbp]
\centering
\includegraphics[width=2.9cm,height=2.2cm]{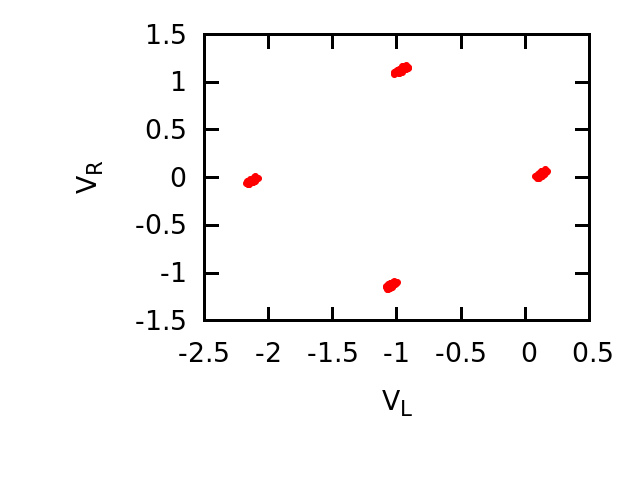}
\includegraphics[width=2.9cm,height=2.2cm]{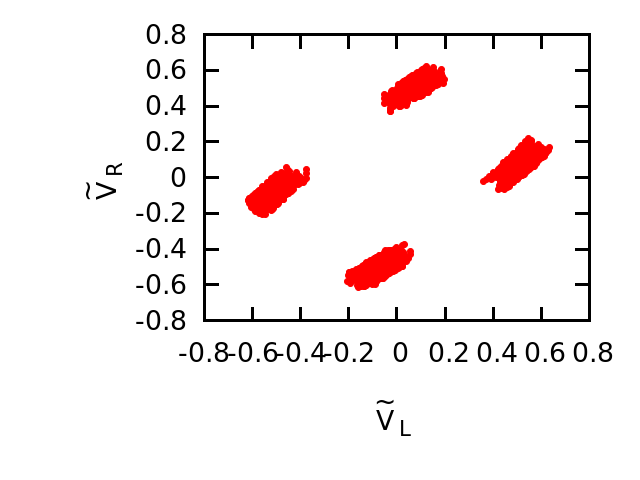}
\includegraphics[width=2.9cm,height=2.2cm]{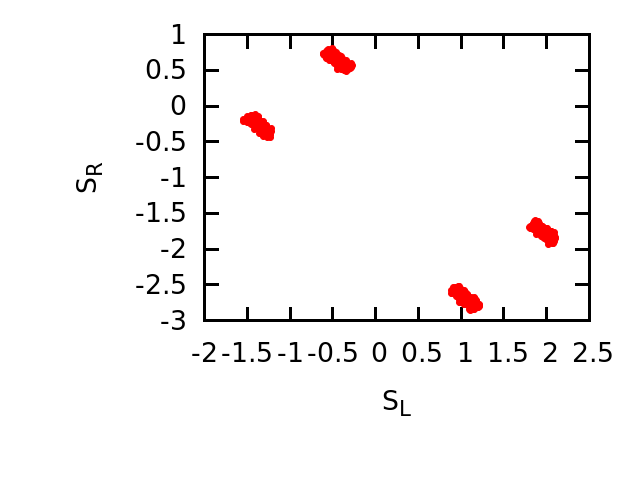}
\includegraphics[width=2.9cm,height=2.2cm]{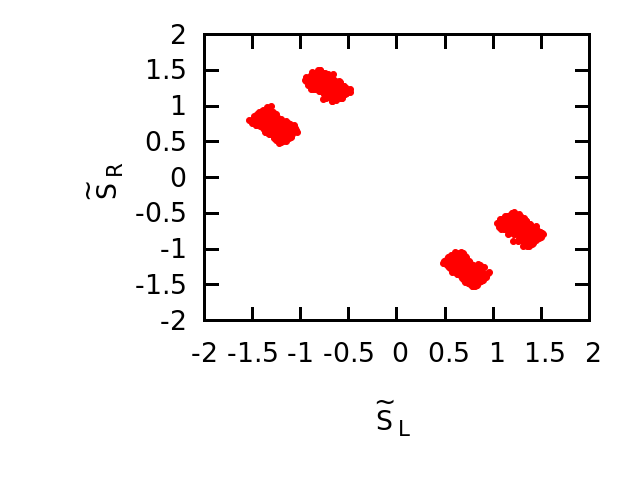}

\includegraphics[width=2.9cm,height=2.2cm]{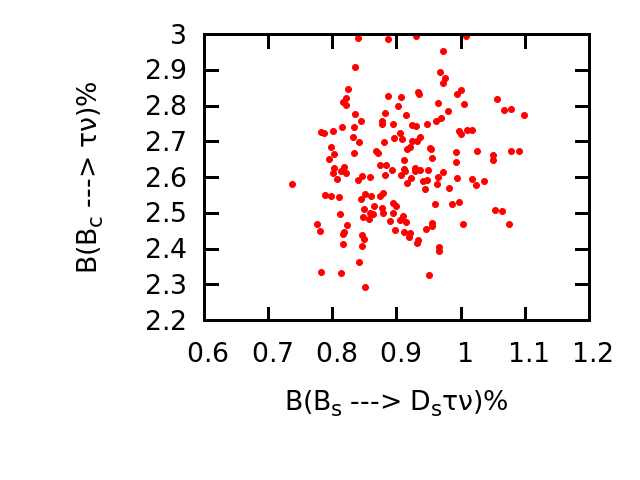}
\includegraphics[width=2.9cm,height=2.2cm]{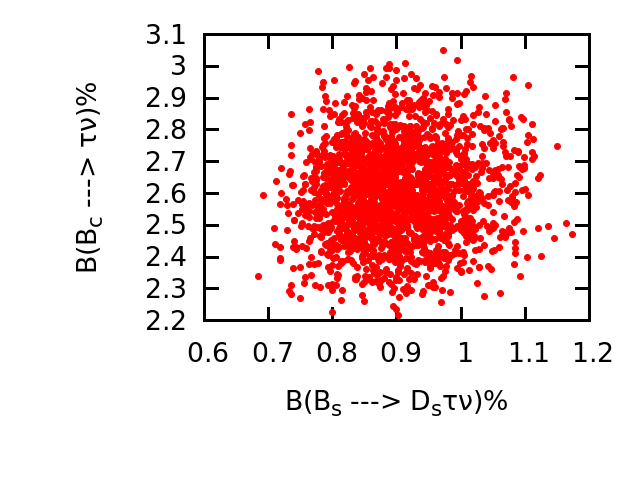}
\includegraphics[width=2.9cm,height=2.2cm]{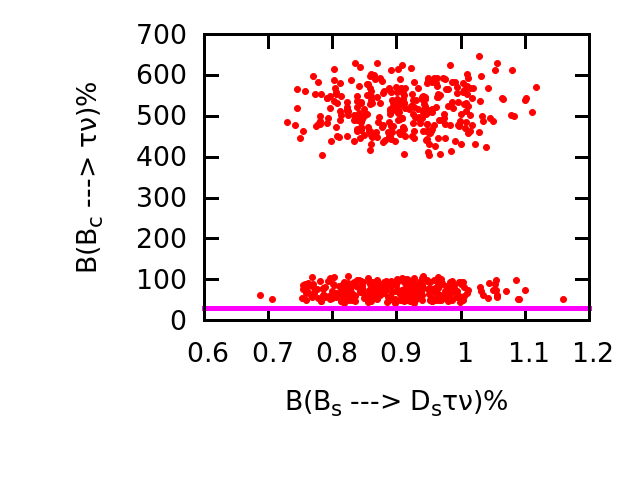}
\includegraphics[width=2.9cm,height=2.2cm]{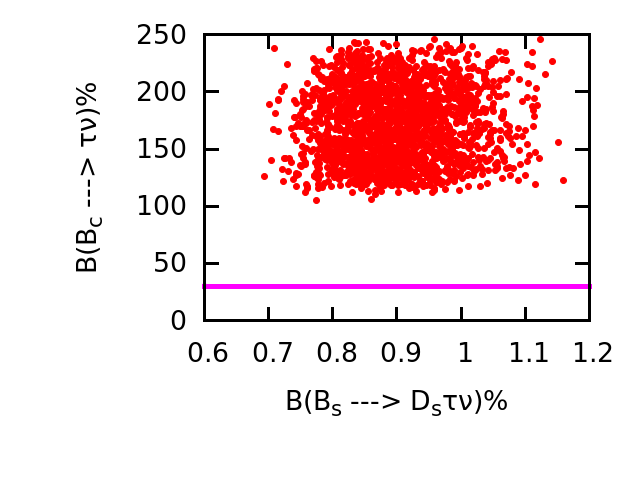}
\caption{{Allowed ranges of $(V_L$, $V_R)$ , $(\widetilde{V}_L, \widetilde{V}_R)$, $(S_L, S_R)$ and $(\widetilde{S}_L, \widetilde{S}_R)$
NP couplings are shown in the top panel once $1\sigma$ constraint coming from the measured values of $R_D$ and $R_{D^{\ast}}$ are imposed. 
We show in the bottom panel the allowed ranges in $\mathcal B(B_c \to \tau\nu)$ and $\mathcal B(B_s \to D_s\tau\nu)$ in the presence of 
respective NP couplings.}}
\label{s1}
\end{figure}

\begin{table}[htbp]
\centering
\begin{tabular}{|c|c|c||c|c|c|}
\hline
Observables &Central value &Range & Observables &Central value &Range\\
\hline
\hline
$\mathcal B_e\%$ &$2.238$ &$[2.013, 2.468]$&$\mathcal B_{\tau}\%$ &$0.670$ &$[0.619, 0.724]$\\
\hline
$P^e$ &$-1.00$ &$-1.00$&$P^{\tau}$ &$0.320$&$[0.273, 0.365]$\\
\hline
$A^e_{FB}$ &$0.00$ &$0.00$&$A^{\tau}_{FB}$ &$0.360$ &$[0.356, 0.363]$\\
\hline
$C^e_F$ &$-1.5$ &$-1.50$&$C^{\tau}_F$ &$-0.271$ &$[-0.253, -0.289]$\\
\hline
\end{tabular}
\caption{SM prediction of various observables for the $e$ and the $\tau$ modes}
\label{tabsm}
\end{table}

\begin{figure}[htbp]
\centering
\includegraphics[width=2.9cm,height=2.2cm]{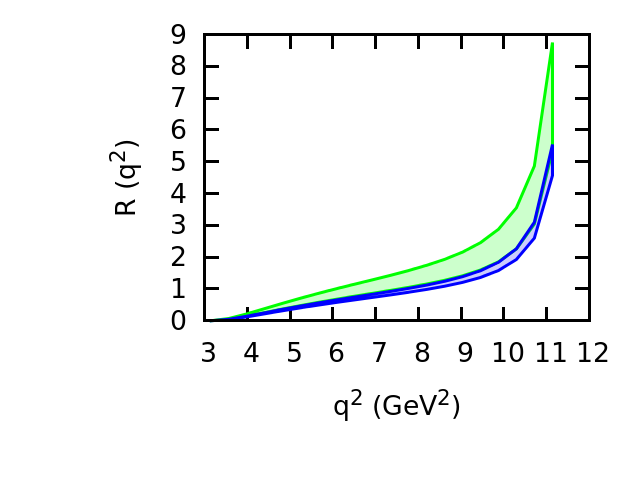}
\includegraphics[width=2.9cm,height=2.2cm]{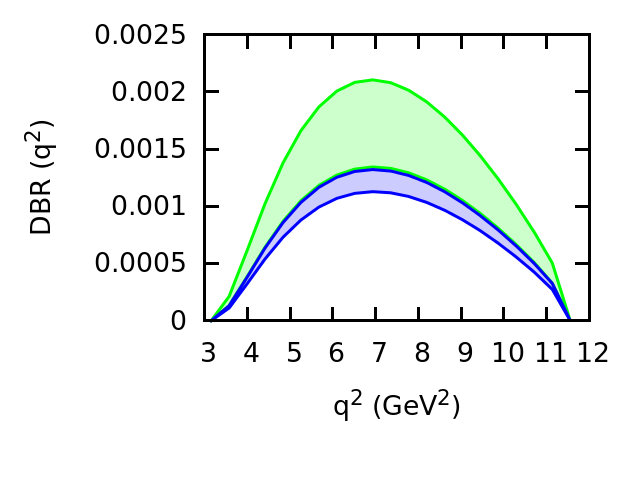}
\includegraphics[width=2.9cm,height=2.2cm]{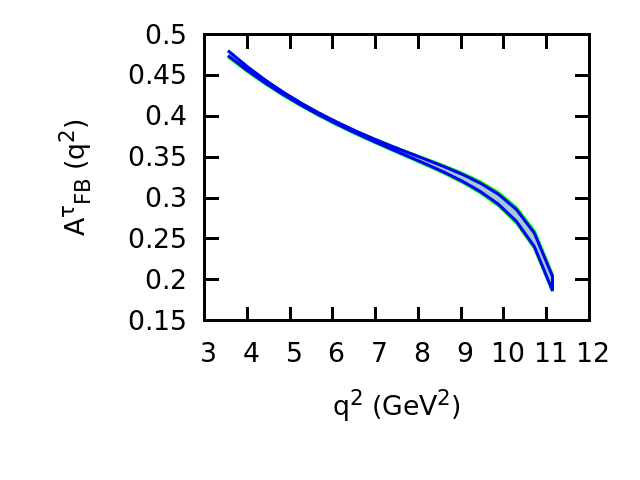}
\includegraphics[width=2.9cm,height=2.2cm]{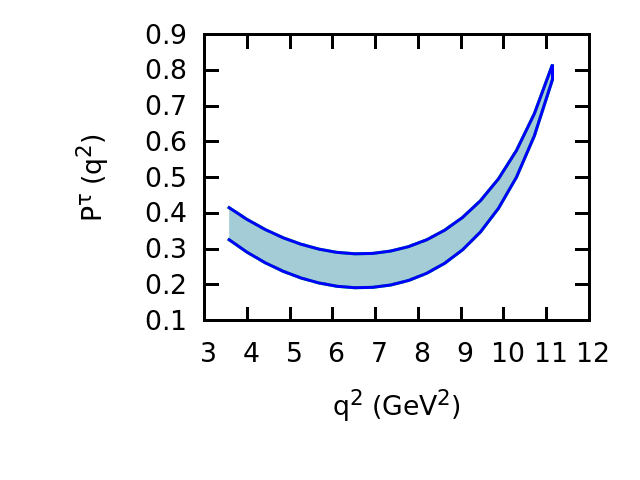}

\includegraphics[width=2.9cm,height=2.2cm]{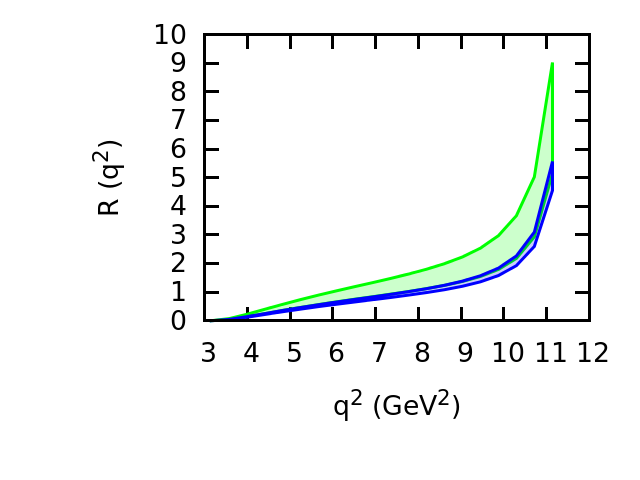}
\includegraphics[width=2.9cm,height=2.2cm]{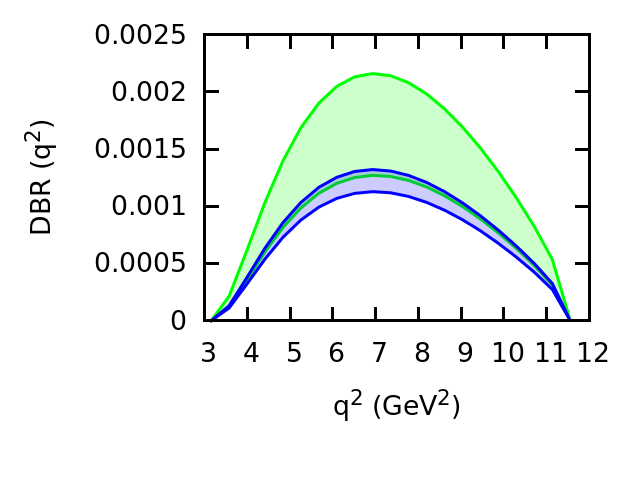}
\includegraphics[width=2.9cm,height=2.2cm]{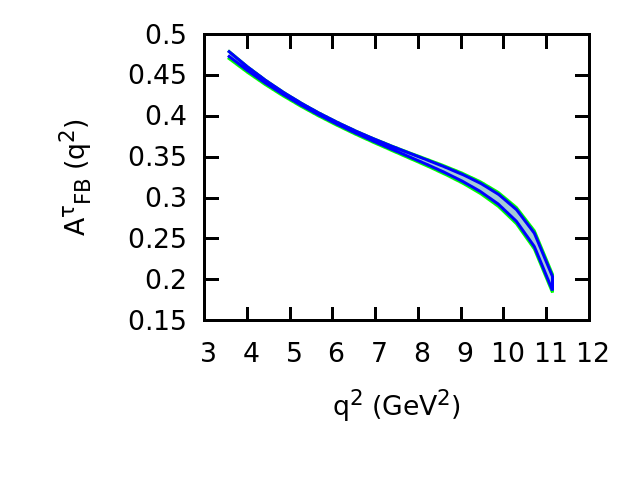}
\includegraphics[width=2.9cm,height=2.2cm]{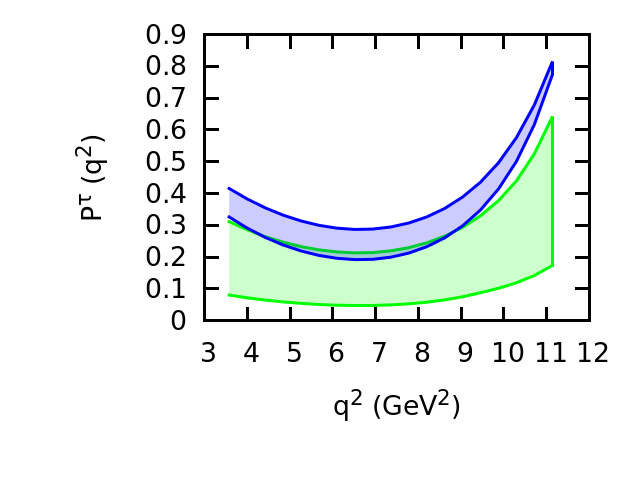}
\caption{{$R(q^2)$, $DBR (q^2)$, $A_{FB}^{\tau} (q^2)$, $P^{\tau} (q^2)$ using $(V_L$, $V_R)$ (top) and $(\widetilde{V}_L, \widetilde{V}_R)$ (bottom) 
NP couplings (green band). The corresponding $1\sigma$ SM range is shown with the blue band.}}
\label{s2}
\end{figure}

We now proceed to discuss the various NP effects in $B_s \to D_s \tau \nu$ decay mode.
We consider four different NP scenarios each containing two NP couplings at a time: $(V_L, V_R)$, $(S_L, S_R)$, $(\widetilde{V}_L, \widetilde{V}_R)$ 
and $(\widetilde{S}_L, \widetilde{S}_R)$. 
To get the allowed NP parameter space we impose the $1\sigma$ constraints coming from the measured ratio of branching ratio $R_D$ and $R_{D^*}$
as well as the requirement of $\mathcal{B}(B_c \to \tau \nu)\le 30\%$ from the LEP data~\cite{Akeroyd:2017mhr}. 
In Fig.~\ref{s1} we show the allowed ranges of each NP couplings and the corresponding $\mathcal{B}(B_s \to D_s \tau \nu)$ and 
$\mathcal{B}(B_c \to \tau \nu)$ allowed regions.
It is observed that the $\mathcal{B}(B_c \to \tau \nu)$ put
a severe constraint on the scalar NP couplings and 
hence in our present analysis we omit the related discussions. 
Table~\ref{t1} reports the allowed ranges of each observables for $B_s \to D_s \tau \nu$ decay mode when $(V_L, V_R)$ and 
$(\widetilde{V}_L, \widetilde{V}_R)$ NP couplings are present. 
We show the $q^2$ dependency of $R(q^2)$, $DBR (q^2)$, $A_{FB}^{\tau} (q^2)$, $P^{\tau} (q^2)$ using $(V_L$, $V_R)$ (top) and $(\widetilde{V}_L, \widetilde{V}_R)$ (bottom) 
NP couplings in Fig.~\ref{s2}. 
In the presence of $(V_L, V_R)$ NP couplings, we observe that $DBR (q^2)$ and $R(q^2)$ deviate considerably from the SM expectation whereas, no deviations are found in 
$A_{FB}^{\tau} (q^2)$, $P^{\tau} (q^2)$, $C_{F}^{\tau} (q^2)$. 
Similarly, in the presence of $(\widetilde{V}_L, \widetilde{V}_R)$ NP couplings the deviation is observed in $P^{\tau} (q^2)$ along with $DBR (q^2)$ and $R(q^2)$. 
Hence the polarization fraction of the charged lepton $P^{\tau} (q^2)$ can be used to distinguish between these two scenarios. 

\begin{table}[htbp]
\centering
\begin{tabular}{|c|c|c|c|c|c|}
\hline
 & $\mathcal B \%$ & $R_{D_s}$ & $P^{\tau}$ & $A^{\tau}_{FB}$ & $C^{\tau}_F$ \\
\hline
\hline
$(V_L$, $V_R)$ & $[0.733, 1.115]$ & $[0.329, 0.496]$ & $[0.234, 0.403]$ & $[0.352, 0.364]$ & $[-0.239, -0.305]$ \\
\hline
$(\widetilde{V}_L, \widetilde{V}_R)$ & $[0.684, 1.174]$ & $[0.307, 0.519]$ & $[0.064, 0.276]$ & $[0.356, 0.363]$ & $[-0.253, -0.289]$ \\
\hline
\end{tabular}
\caption{Allowed ranges of various observables with $(V_L$, $V_R)$ and $(\widetilde{V}_L, \widetilde{V}_R)$ NP couplings.}
\label{t1}
\end{table}

\section{Conclusion}
Based on the anomalies present in $R_D$ and $R_{D^*}$, we study their implication on $B_s \to D_s \tau \nu$ decay mode within the SM and within various NP scenarios. 
We find that only vector type NP couplings satisfy the $\mathcal{B}(B_c \to \tau \nu)$ constraint
whereas, the scalar type NP couplings are ruled out.
Hence, studying the $B_s \to D_s \tau \nu$ decay mode will serve as an important stepping stone for $R_D$ and $R_{D^*}$ anomalies.

\end{document}